\documentclass[notitlepage,twocolumn,prl,tightenlines,superscriptaddress,showpacs,floatfix]{revtex4-1} 

\usepackage{amsmath,amssymb,amsfonts,latexsym}
\usepackage[colorlinks=true, linkcolor=blue, citecolor=blue]{hyperref}
\usepackage{bbm}
\usepackage[mathcal]{euscript}
\usepackage{graphicx}
\usepackage{epsfig}
\usepackage{color}

\usepackage{psfrag}
\usepackage{txfonts,pxfonts,wasysym,}
\usepackage{pifont}
\usepackage[caption = false]{subfig}
\newcommand{\gul}{Gulliver UMR CNRS 7083, ESPCI Paris, PSL Research University, 10 rue Vauquelin, 75005 Paris, France}
\newcommand{\tav}{School of Physics and Astronomy, and the Center for Physics and Chemistry of Living Systems, Tel Aviv University, Tel Aviv 6997801, Israel}

\newcommand{\be}{\begin{equation}}
\newcommand{\ee}{\end{equation}}
\newcommand{\ben}{\begin{equation*}}
\newcommand{\een}{\end{equation*}}
\newcommand{\ba}{\begin{eqnarray}}
\newcommand{\ea}{\end{eqnarray}}

\begin{document}
\graphicspath{{./Figures/}}

\title{Reconfiguration, Interrupted Aging and Enhanced Dynamics of a Colloidal Gel using Photo-Switchable Active Doping}

\author{Mengshi Wei}
\affiliation{\gul}
\author{Matan Yah Ben Zion}
\affiliation{\tav}
\author{Olivier Dauchot}
\affiliation{\gul}

\date{\today}

\begin{abstract}
We study light-activated quasi-2d gels made of  a colloidal network doped with Janus particles. Following the gel formation, the internal dynamics of the gel are monitored before, during, and after the light activation. We monitor both the structure and dynamics, before, during and after the illumination period. The mobility of the passive particles exhibits a characteristic scale-dependent response. Immediately following light activation, the gel displays large-scale reorganization, followed by progressive, short-scale displacements throughout the activation period. Albeit subtle structural changes (including pore opening and widening and shortening of strands) the colloidal network remains connected, and the gel maintains its structural integrity. Once activity is switched off, the gel keeps the memory of the structure inherited from the active phase. Remarkably, the motility remains larger than that of the gel, before the active period. The system has turned into a genuinely different gel, with frozen dynamics, but with more space for thermal fluctuations. The above conclusions remain valid long after the activity period.
\end{abstract}

\maketitle
Programmable self-assembling is a widely used bottom-up approach for achieving materials with desired properties~\cite{whitesides2002self, sacanna2012magnetic, ben2017self, kennedy2022self}. A common route for self-assembly consist in engineering the inter-colloid potential such that the structure, be it equilibrium or kinetically arrested, matches the expected design. 
The advent of active Janus colloids has created the opportunity to engineer at the local scale, not only the interactions, but also the dynamics of the self-assembling process. Such active doping has been shown both experimentally and numerically to be a realistic strategy for either driving the system toward its thermodynamically favored crystalline target structure~\cite{ni2014crystallizing, dietrich2018active, mallory2020universal}, or modulating the structure of isotropic colloidal gels and glasses~\cite{van2016fabricating, singh2017non, omar2018swimming, ramananarivo2019activity, janssen2019active}.

The case of gels is of particular interest. First there are strong experimental evidences of anomalous mechanical responses and original dynamics in biological and reconstituted biopolymer networks~\cite{koenderink2009active, sanchez2011cilia, sanchez2012spontaneous, berezney2022extensile}. These observations have in turn driven a large amount of theoretical work aiming at deciphering the specific microscopic mechanisms and formulating an effective medium theory~\cite{gardel2004elastic, storm2005nonlinear, mizuno2007nonequilibrium, ronceray2016fiber, goldstein2019stress}. Second, from a rather fundamental perspective, gels are disordered out of equilibrium materials with plethora of metastable configurations. 
This offers new opportunities for exploring the configuration landscape and equilibration dynamics in the presence of active dopants. Yet in a biological gel activity and elasticity are intertwined, and reciprocate intermittently throughout the formation cycle of the gel, making it extremely challenging to decipher the respective contributions of elasticity and activity.

We concentrate on the a priori simpler case of colloidal gels, for which a few results have been obtained so far. It was shown experimentally that a fractal cluster colloidal gel with embedded active Janus colloids displays enhanced dynamics and a reduction in linear viscoelastic moduli in proportion to activity, while its yield stress decreases significantly even for a very small fraction of dopant~\cite{szakasits2017dynamics, szakasits2019rheological, saud2021yield}. The numerical study of the coarsening dynamics of a model colloidal gel former, including active particles, lead to the prediction of a phase diagram parametrized by the intensity and the directional persistence of the active forces~\cite{omar2018swimming}.  When the active forces are smaller, but comparable to the adhesion forces, the coarsening dynamics accelerates as activity helps drive the coalescence of the gel strands, with a strong amplification of the effect with increasing persistence.  When the active forces are larger than the adhesion ones, the active colloids are no longer bound to other particles and start mediating new effective interactions amongst the passive particles. In the above studies the active doping is present from the early stage of gelation and therefore biases the whole gelation process, leaving aside two important questions: (i) what is the response of a passive gel with embedded Janus, yet not activated, particles to a stepwise switch of the activity? (ii) what is the fate of the gel, once the activity is switch off, after a given period of activity? This last question is of particular interest if one is to imprint permanent functionalities in a gel using active doping.

In this letter, we design and study of a quasi 2D colloid gel composed of a mixture of passive and Janus particles the activity of which is switched on using light \emph{once the gel is formed} and kept constant for a while before it is switched off. The activity level is tuned by the light intensity and kept such that the active colloids remain bounded to the gel. We monitor both the structure and dynamics, before, during and after the illumination period. We find that during the active phase, the mobility of the passive particles not only increases with activity, but also exhibits a characteristic scale-dependent response to activity.  Our main findings concern the reconfigured gel, once activity is turned off.
First, the gel keeps memory of the structure inherited from the active phase. Second, it exhibits a \emph{larger} mobility than that of the gel before the active period. This increase of the dynamics correlates well with an increase of structural heterogeneities in the gel. Altogether, the system has turned into a genuine different gel, and remains unaltered during several weeks.

\begin{figure}[t!]
\vspace{-0mm}
\includegraphics[width=0.95\columnwidth,trim = 0mm 0mm 0mm 0mm, clip]{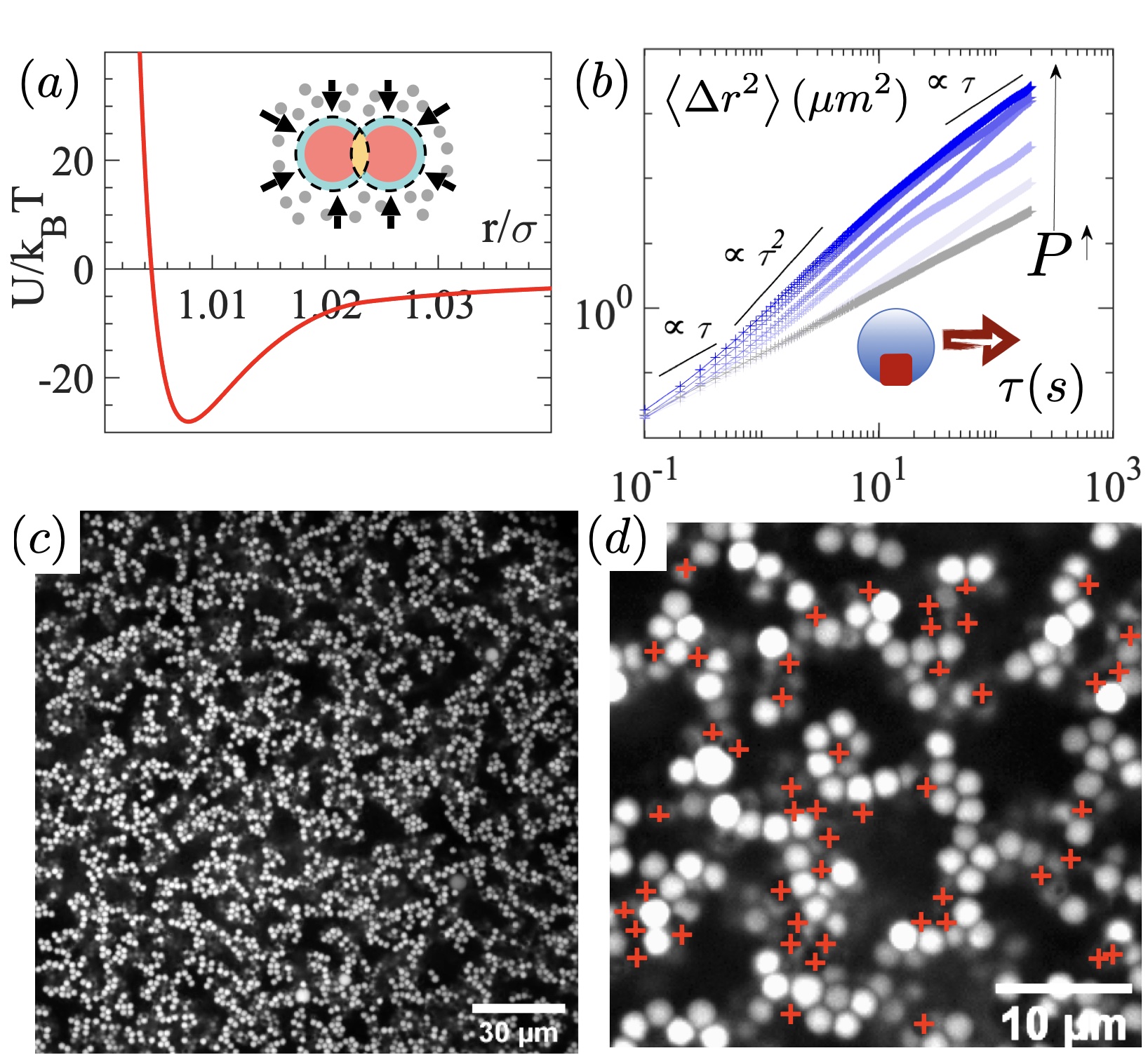}
\vspace{-0mm}
\caption{{\bf Experimental system:} (a) Attraction potential between the colloids composing the gel, as obtained by depletion mediated by PEO polymer. (b) Mean square displacement (MSD) of the janus particles suspended at very low density in the actual solvent used when preparing the gel, activated with different blue light intensities P. (c) Quasi 2D gel obtained from the joint sedimentation of polystyrene colloids and janus particles in the depleting and swimming buffer. (d) Confocal image of the quasi-2D gel, with the janus particles marked with a red cross.}
\label{fig:system}
\vspace{-5mm}
\end{figure}

The colloidal gel is obtained by the sedimentation of a mixture of passive $\sigma = 2 \mu {\rm m}$ diameter polystyrene particles and active $1.5 \mu {\rm m}$ diameter particles that consist in a hematite $(\alpha-Fe_2O_3)$ cube partially protuding outside a shell of 3-methacryloxypropyltrimethoxysilane (TPM)~\cite{palacci2013living}, in the presence of polyethylene glycol (PEO) acting as a depletant agent. One serious challenge in designing such a system is to compromise between the need for screening the electrostatic repulsion between the colloids, while keeping a strong enough motility for the active particles, the later being known to decrease in the presence of salt. All experimental parameters are provided in the Supp. Materials. The resulting interaction potential, estimated from the addition of the Asakura-Oosawa~\cite{asakura1954interaction, vrij1977polymers_depletion_vrij} and DLVO~\cite{derjaguin1993theory, derjaguin1941theory, verwey1947theory}potentials, is displayed on Fig.~\ref{fig:system}(a). The mobility of the active particles in the presence of blue light ($\lambda = 494 /40$ nm) is characterized by their mean square displacements (MSD) (Fig.~\ref{fig:system}(b)), as measured in a dilute suspension of the active particles alone, in the buffer solution including the PEO, which increases the drag coefficient $\zeta$. Assuming a description of the active particles in terms of standard Active Brownian Particles~\cite{franke1990galvanotaxis, howse2007self}, we infer from these measurements that the persistent time of the active particle $\tau_p =16.5 \pm 7.5 $s, independently of the light intensity, while the swimming speed $U_0$ increases up to $0.57 \pm 0.13 \mu {\rm m}/{\rm s}$ when the light intensity $P$ reaches $154 {\rm mW/mm}^2$ (see also Fig S1, Supp. Mat.).  

\begin{figure}[t!]
\vspace{-0mm}
\includegraphics[width=0.95\columnwidth,trim = 0mm 0mm 0mm 0mm, clip]{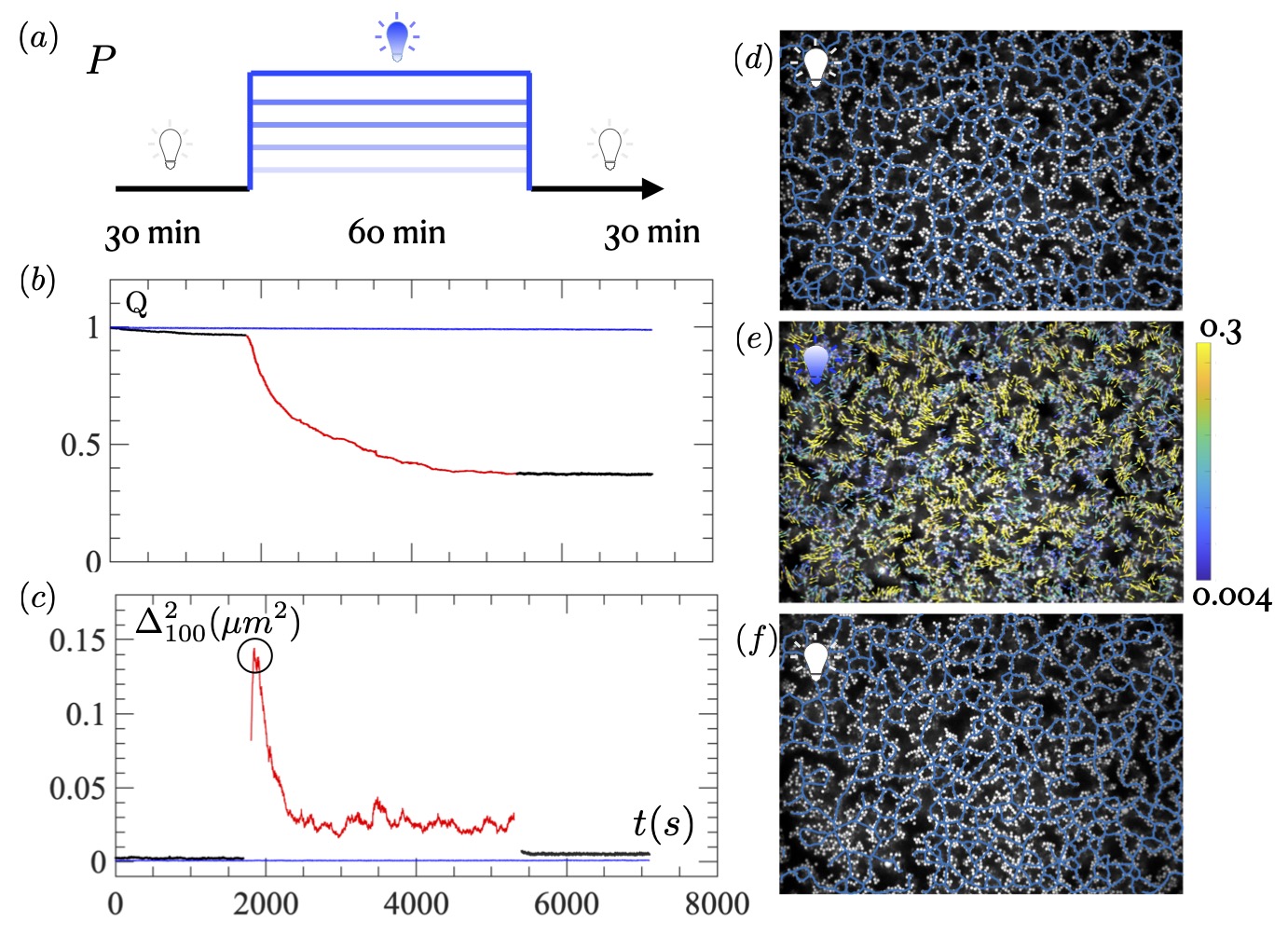}
\vspace{-0mm}
\caption{{\bf Gel reconfiguration during activation :} (a) Illumination protocole : 24 hours after the end of sedimentation, we start recording $30$ min of the dynamics of a metastable fractal gel, followed by $60$ min of activation with blue light of various intensities and $30$ min of relaxation in the absence of blue light. (b) The overlap $Q(t)$, as defined in the main text, between the gel configuration at time $t$ and that at time $t=0$. (c)  $\Delta^2_{100}(t)$, square displacements computed over $\tau=100$ s, averaged over the passive colloids. (d) Gel configuration before light illumination (skeleton overlaid in blue). (e) Displacement field of the passive colloids computed over $\tau=100$ s, around the peak of $\Delta^2_{100}$, as indicated by the black circle in panel (c), color coded by their magnitude. $P =118 {\rm mW/mm}^2$.}
\label{fig:protocole} 
\vspace{-5mm}
\end{figure}

After sedimentation, the deposited gel is left to relax for $24$ hours before observation. A typical illumination protocol (Fig.~\ref{fig:protocole}-a) consists of $30$ min recording of the gel in the absence of blue light, followed by $60$ min of blue light illumination, the intensity $P$ of which can be varied, and yet $30$ min in the absence of blue light. The gel is monitored under red light ($\lambda = 647$ nm) using a confocal microscope equipped with a 40x oil objective and a CDD camera, resulting in images of size $249 x 173 \mu m^2$, with a spatial resolution of 0.1625 $\mu {\rm m}$/pixel, acquired at $1$ frame per second. Using standard image processing techniques, we track the trajectories ${\bf r}_i (t)$ of the passive particles, from which we compute the displacement field $\Delta {\bf r}_i(t,\tau)  = {\bf r}_i(t+\tau) - {\bf r}_i(t)$ , the overlap $Q(t)$ and the square displacement averaged over the passive particles $\Delta^2_{\tau} (t)$~:
\begin{eqnarray}
Q(t) &=\frac{1}{N} \sum_i \exp(-\frac{\Delta {\bf r}_i^2(0,t)}{a^2})\\
\Delta^2_{\tau} (t) &= \frac{1}{N} \sum_i \Delta {\bf r}_i^2(t,t+\tau), 
\end{eqnarray}
where $N\simeq 2500$ is the number of tracked passive particles and $a=\sigma/3$ is a characteristic scale of motion.

During the first passive period, the gel exhibits a slow dynamics with $\Delta^2_{100}\simeq 2.10^{-3}\mu {\rm m}^2$, together with a slow aging, attested by the decay of $Q(t)$ (Fig.~\ref{fig:protocole}-b). When light activation is switched on ($P = 118 {\rm mW/mm}^2$), a fast response of the dynamics, followed by a rapid relaxation, leads to a peak of $\Delta^2_{100}\simeq 150.10^{-3}\mu {\rm m}^2$, before the dynamics settles to weaker values of $\Delta^2_{100}\simeq 20.10^{-3}\mu {\rm m}^2$, still significantly larger than in the absence of light. Associated with this dynamical response, the gel reorganizes, as demonstrated by the sharp decay of $Q$ and illustrated on Figures~\ref{fig:protocole}-(b), with the displacement field $\Delta {\bf r}_i(t,100)$, computed around the peak of $\Delta^2_{100}$. When activation is switched off, we observe that (i) $Q$ remains constant: the gel is frozen and keeps memory of the reorganization that took place during activation; and (ii) $\Delta^2_{100}\simeq 5.10^{-3}\mu {\rm m}^2$ is larger than before activation: the gel is more motile. Figures~\ref{fig:protocole}-(d,f) display typical gel configurations before and after illumination. The structure of the gel is certainly modified, but retains its overall integrity (see overlaid skeleton Fig S3 in Supp. Mat.)

\begin{figure}[t!]
\vspace{-0mm}
\includegraphics[width=0.95\columnwidth,trim = 0mm 0mm 0mm 0mm, clip]{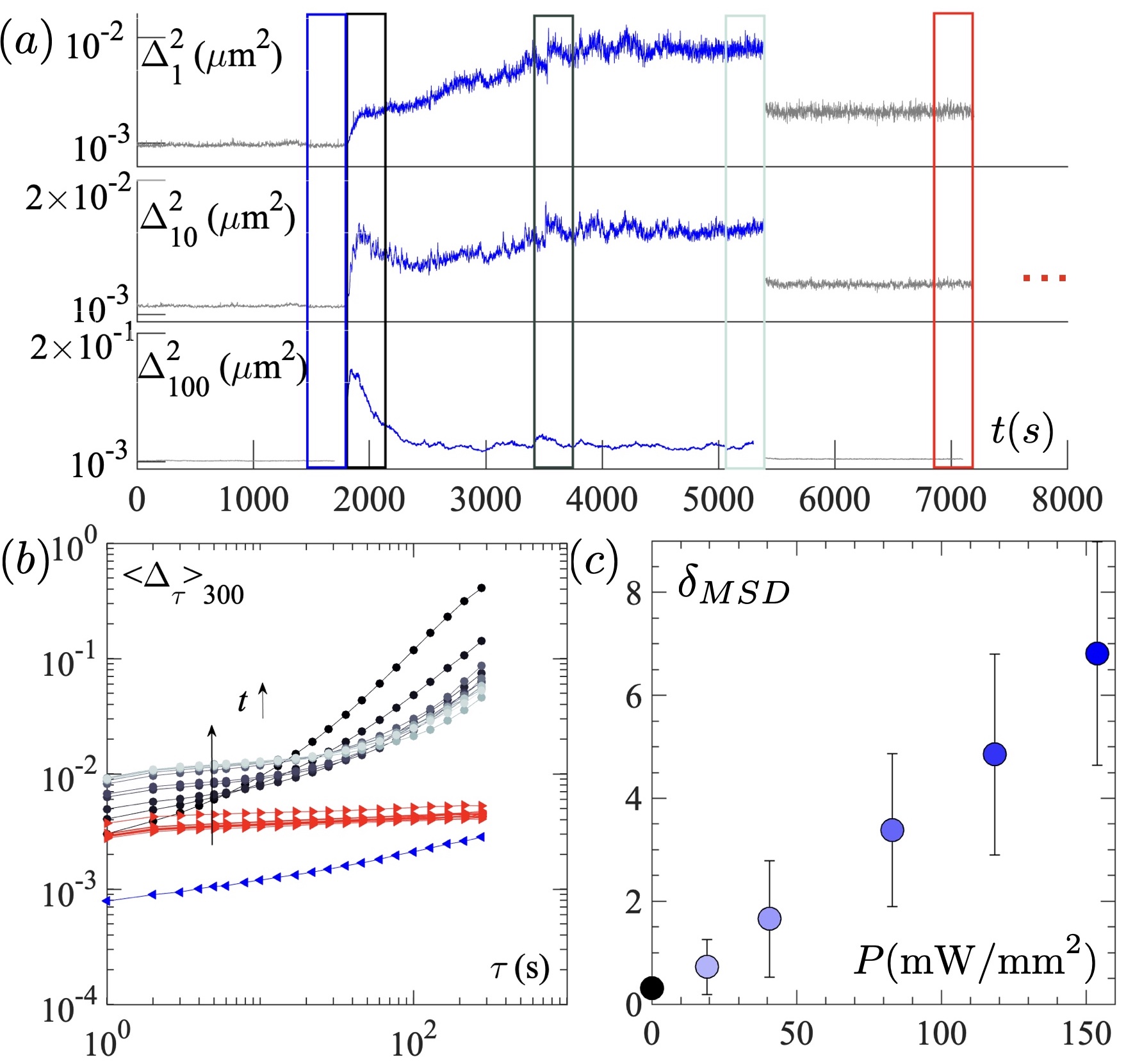}
\vspace{-0mm}
\caption{{\bf Averaged dynamics:} (a) Averaged square displacements $\Delta^2_{1}(t), \Delta^2_{10}(t), \Delta^2_{100}(t)$; (b) MSD $\left<\Delta^2_\tau\right>_{300}$, computed at successive times indicated in panel (a) before (blue), during (gray levels) and after (red) activation; the numerous curves in red correspond to the same measurement repeated every two hours after activation, during 30 hours ($P = 118{\rm mW/mm}^2$); (c) $\delta_{msd}$, the relative increase of the short time ($\tau = 10$ s) MSD after activation, as a function of the light intensity during the activation period.$P$.}
\label{fig:avg_dyn}
\vspace{-7mm}
\end{figure}

The dynamics of the gel strongly depend on the timescale $\tau$ on which it is probed (Fig.~\ref{fig:avg_dyn}-a). While on long time scales, $\tau=100$, the response takes the form of a sharp peak immediately following the activation, on short time scales, $\tau = 1$, the displacements essentially increase continuously during the whole activation period.  As a result, the MSD, $\left<\Delta^2_\tau\right>_{300}$, averaged over successive time windows of duration $300$s also strongly depends on the time $t$ at which it is probed. Before activation (Fig.~\ref{fig:avg_dyn}-b, blue curve), the dynamics is strongly sub-diffusive as expected for a slowly aging gel. Just after activation, (Fig.~\ref{fig:avg_dyn}-b, black curve) the dynamics is diffusive as if the gel had melted. Later during the activation period, the gel recovers a sub-diffusive dynamics although with a significantly larger amplitude (Fig.~\ref{fig:avg_dyn}-b, gray curves). The dynamical response therefore exhibits a characteristic crossover for a timescale $\tau^* \simeq 10$ s, corresponding to displacements of the order of $\Delta^* \simeq 10^{-1} \mu {\rm m}$. This crossover suggests that the large scale displacements $\Vert \Delta {\bf r}_i \Vert > \Delta^*$ triggered just after activation, contribute to a reorganization of the gel allowing for increasingly larger small scale displacements $\Vert\Delta {\bf r}_i\Vert < \Delta^*$ at all times. The gel after activation keeps memory of this reorganization for a long time as demonstrated by the amplitude of $\left<\Delta^2_\tau\right>_{300}$ after activation, which remains larger than that before activation for whatever time we waited ($30$ hours on Fig.~\ref{fig:avg_dyn}-b, red curves, $11$ days on Fig. S5, Supp. Mat.). We also stress that $\left<\Delta^2_\tau\right>_{300}$ after activation is flat, indicating a frozen system with fully caged particles. 

\begin{figure}[t!]
\vspace{-0mm}
\includegraphics[width=0.95\columnwidth,trim = 0mm 0mm 0mm 0mm, clip]{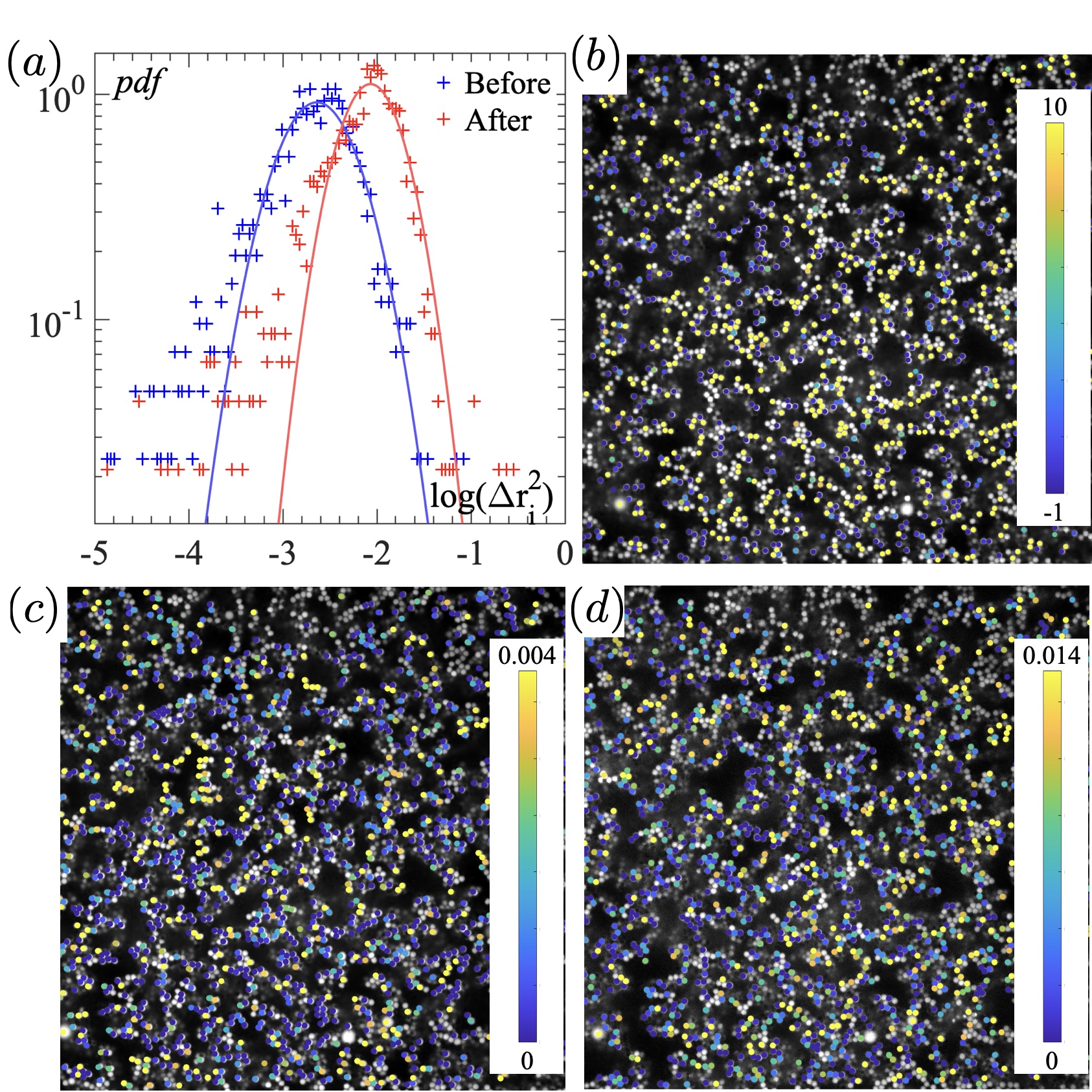}
\vspace{-0mm}
\caption{{\bf Local dynamics:} (a) Probability distribution of the individual square displacements $\Delta {\bf r}_i^2 (t,10)$ before and after activation. Continuous lines indicate gaussian distributions for comparison. (b) Spatial distribution of $\delta_{\rm MSD}$. (c, d) Spatial distribution of $\Delta {\bf r}_i^2 (t,100)$ respectively before and after activation. The color bar ranges are scaled to 2 times the average value.}
\label{fig:loc_dyn}
\vspace{-5mm}
\end{figure}

Altogether the activation interrupted the slow aging of the gel and rapidly drove it into a new frozen state, which exhibits larger thermal fluctuations! This unexpected result has been confirmed, repeating the experiment over more than $80$ gels, prepared in the same way, and activated with varying light intensities. The relative increase of the MSD $\delta_{\rm MSD}  = (\left<\Delta_\tau\right>_{>} - \left<\Delta_\tau\right>_{<})/\left<\Delta_\tau\right>_{<}$, where $\left<\Delta_\tau\right>_{<}$ and $\left<\Delta_\tau\right>_{>}$ respectively denote the mean square displacement computed before and after activation, is a clear growing function of the light intensity (Fig.~\ref{fig:avg_dyn}-c). We have also checked that the above phenomenology does not rely on an alteration of the subtract nor of the PEO by the illumination (see Fig S4, S5, Supp. Mat.)

Figure~\ref{fig:loc_dyn} demonstrates that the above phenomenology homogeneously takes place in the gel. The distributions of $\Delta  {\bf r}_i^2(t,10)$, before and after activation are very similar and essentially simply shifted to larger values. Their spatial distribution is homogeneous, both before and after activation, even a bit more homogeneous following activation. Finally $\delta_{\rm MSD}$, when  computed for individual particles is also homogeneously distributed in space. 

\begin{figure}[t!]
\vspace{-0mm}
\includegraphics[width=0.95\columnwidth,trim = 0mm 0mm 0mm 0mm, clip]{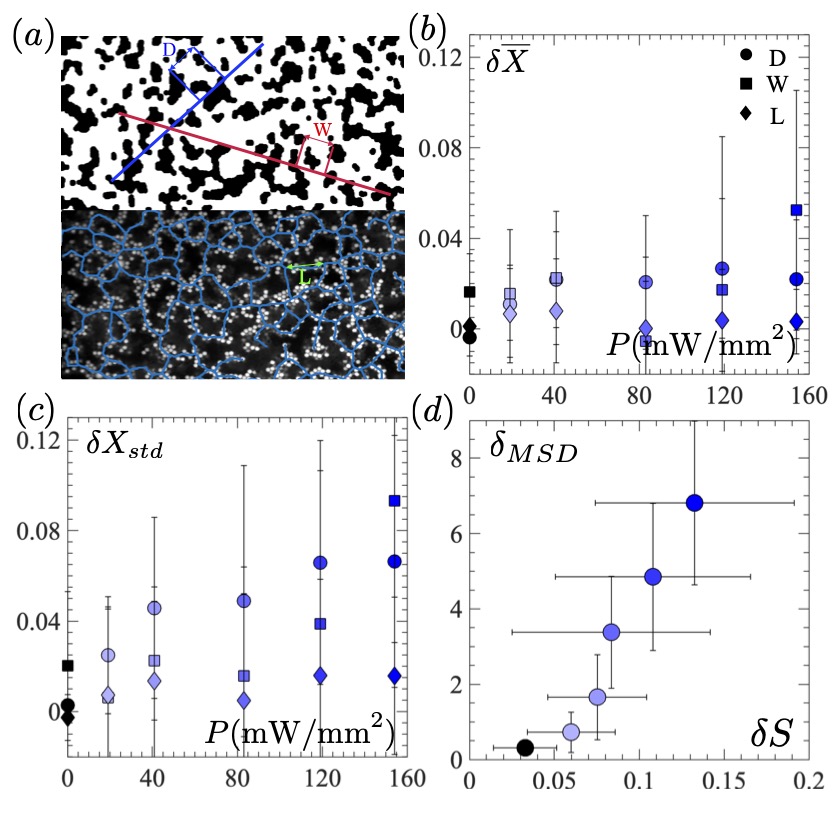}
\vspace{-0mm}
\caption{{\bf Gel structure} : (a) Top: Binary density field and a sample of chord lengths defining the strands width $W$ and the pores size $D$; Bottom: Gel skeleton as obtained by morphological analysis of the binary density field (see Supp. Mat.); the green arrow indicates the length $L$ of a gel strand; (b) Relative variation of the mean of the pores size  $\delta\overline{D}$, the strand width $\delta\overline{W}$, and the  link length $\delta{\overline{L}}$,  and (c) relative variation of their standard deviation  $\delta D_{std}$ ,  $\delta W_{std}$, and $\delta L_{std}$ as a function of the light intensity. (d) Correlation between the relative increase of the short time mean square displacement $\delta_{msd}$ and the structural change $\delta S$ as defined in the text.}
\label{fig:struct}
\vspace{-5mm}
\end{figure}

As stated above, there is no obvious direct visual evidence of a modification of the structure, which could simply explain the observed dynamical changes. The pair correlation functions after activation has slightly larger first and second peaks, indicating a more compact structure (see Fig S2, Supp. Mat.). We further characterize the gel structure by performing a morphological image analysis, which consists in approximating the instantaneous density field by a binary image, from which we extract a set $Xq$ of structural parameters: the width of the gel strands, $X1 = W$, and the size of the gel pores, $X2 = D$. We further skeletonize the binary image, to extract the strands lengths, $X3 = L$ (Fig.~\ref{fig:struct}-a) and Supp. Mat.). We then compute their spatio-temporal average $\overline Xq$ and standard deviation $Xq_{std}$ before ($<$) and after ($>$) activation, for each gel. We denote $\delta \overline Xq  = (\overline Xq_{>} - \overline Xq_{<})/ \overline Xq_{<}$ and $\delta Xq_{std}  = (Xq_{std,>} - Xq_{std,<})/ Xq_{std,<}$ the variation of these statistical descriptors of the gel structure across activation.  The average strand width and the pore size increase by a few percent, the variation being larger with stronger illumination (Fig.~\ref{fig:struct}-b), while the strand length remains essentially unchanged. These admittedly small variations are compatible with the evolution of the pair correlation function and suggest a weak coarsening of the gel structure. The evolution of the gel structure appears more clearly on the standard deviations, which increase more systematically, indicating a more heterogeneous structure after activation (Fig.~\ref{fig:struct}-c). Defining $\delta S = \left( \sum_q \delta Xq_{std}^2 \right)^{1/2}$, a parameter quantifying the overall relative variation of the structure heterogeneity induced by the activation, we observe that the dynamical change correlates very well with it (Fig.~\ref{fig:struct}-d) : the more the structure heterogeneity has increased, the more the short time mean square displacement has increased.

In discussing the above results, it is useful, following~\cite{omar2018swimming}, to compute the strength of the swim force $F_s =  \zeta U_0$ relative to the attractive force scale $F_a = E_0/2R_g$, where $E_0$ is the depth of the attractive potential and $R_g$ is the interaction range given by the gyration radius of the PEO. We find that for the maximal light intensity $F_s/F_a \simeq 10^{-2}$ confirming that the active particles cannot individually break bonds. However, there are also two active energy scales to consider : $E_s = F_s \sigma \simeq 10 k_B T$ and $E_p = F_s l_p \simeq 100 k_B T$, where $l_p = U_0 \tau_p $ is the active persistence length. These energy scale are comparable or larger than $E_0 \simeq 25 k_B T$, indicating that the collective contribution of a few active particles can easily produce local rearrangements. Note that activity not only acts as an effective hot temperature; it can also activate elasticity driven relaxation, which are central to the complex time-dependance signature of aging in gels~\cite{bouzid2017elastically}.

Our observations during the activation period compare well, and extend to higher level of activity, those of Solomon and collaborators~\cite{szakasits2017dynamics, szakasits2019rheological, saud2021yield}: activity increases the dynamics and decreases the gel stiffness, while hardly modifying the structure of the gel. An important difference though, is that we prepare the gel in a passive state, before the response to activity is studied. From that point of view, it is of interest to compare our result to that of the response to a global shear~\cite{masschaele2009direct, masschaele2011flow, hoekstra2003flow}, in the spirit of the theoretical link made between both~\cite{morse2021direct}. In both cases the gel becomes more heterogeneous, with larger voids, locally compacted regions, and rather small changes in the short range structure.  These observations support recent claims about the role of structural heterogeneities in the softness of gels~\cite{del2007length, colombo2013microscopic,rocklin2021elasticity}. More specifically the strong correlation observed between the standard deviation of the elementary structural parameters and the dynamics, could be a first hint towards the existence of a length scale, characterizing the hidden hierarchical structure of the gel, as suggested recently~\cite{bantawa2022hidden}. 

Altogether, we have proposed a new pathway, using switchable local activation, towards a fast synthesis of stable colloidal gels with tunable softness. The present system has a number of tunable parameters such as the packing fraction or the depletion attraction, opening room for optimization.  Extending the present work to 3D gels and investigating their mechanical properties, is a promising route for the future. On the theoretical side,  identifying the type of state the gel has reached, and understanding if and how they differ from their equilibrium counterparts remains a fantastic challenge.

{\it --- Acknowledgments ---}
We thank Laura Rossi, from TU-Delft for providing us with the Hematite-TPM particles.
We thank Patrick C. Royall and  Emanuela del Gado for inspiring discussions. MW acknowledges support from CSC(N°201806120044).
\vspace{-5mm}
\bibliography{Active.bib}


\end{document}